\documentclass[manuscript]{acmart}

\usepackage{algorithm}
\usepackage{algorithmic}
\usepackage{url}
\usepackage{bbm}
\usepackage{bm}
\usepackage{subcaption}
\usepackage[export]{adjustbox}
\usepackage{amsmath}
\usepackage{mathtools}
\usepackage{multirow}
\usepackage{makecell}

\AtBeginDocument{%
  }

\setcopyright{none} 
\copyrightyear{2023}
\acmYear{2023}
\acmDOI{XXXXXXX.XXXXXXX}

\acmConference[RecSys '23]{email}{June 03--05, 2023}{Woodstock, NY}
\acmPrice{15.00}
\acmISBN{978-1-4503-XXXX-X/18/06}

\begin{document}

\title{Knowledge Graph Completion Models are Few-shot Learners: An Empirical Study of Relation Labeling in E-commerce with LLMs}

\author{Jiao Chen}
\authornote{All three authors contributed equally to this research.}
\email{Jiao.Chen0@walmart.com}
\affiliation{%
  \institution{Walmart Global Tech}
  \city{Sunnyvale, CA}
  \country{USA}
}

\author{Luyi Ma}
\authornotemark[1]
\email{luyi.ma@walmart.com}
\affiliation{%
  \institution{Walmart Global Tech}
  \city{Sunnyvale, CA}
  \country{USA}
}

\author{Xiaohan Li}
\authornotemark[1]
\email{xiaohan.li@walmart.com}
\affiliation{%
  \institution{Walmart Global Tech}
  \city{Sunnyvale, CA}
  \country{USA}
}

\author{Nikhil Thakurdesai}
\email{nikhil.thakurdesai@walmart.com}
\affiliation{%
  \institution{Walmart Global Tech}
  \city{Sunnyvale, CA}
  \country{USA}
}

\author{Jianpeng Xu}
\email{jianpeng.xu@walmart.com}
\affiliation{%
  \institution{Walmart Global Tech}
  \city{Sunnyvale, CA}
  \country{USA}
}

\author{Jason H.D. Cho}
\email{Jason.Cho@walmart.com}
\affiliation{%
  \institution{Walmart Global Tech}
  \city{Sunnyvale, CA}
  \country{USA}
}

\author{Kaushiki Nag}
\email{Kaushiki.Nag@walmart.com}
\affiliation{%
  \institution{Walmart Global Tech}
  \city{Sunnyvale, CA}
  \country{USA}
}

\author{Evren Korpeoglu}
\email{EKorpeoglu@walmart.com}
\affiliation{%
  \institution{Walmart Global Tech}
  \city{Sunnyvale, CA}
  \country{USA}
}

\author{Sushant Kumar}
\email{sushant.kumar@walmart.com}
\affiliation{%
  \institution{Walmart Global Tech}
  \city{Sunnyvale, CA}
  \country{USA}
}

\author{Kannan Achan}
\email{kannan.achan@walmart.com}
\affiliation{%
  \institution{Walmart Global Tech}
  \city{Sunnyvale, CA}
  \country{USA}
}

\renewcommand{\shortauthors}{Chen et al.}

\begin{abstract}
    Knowledge Graphs (KGs) play a crucial role in enhancing e-commerce system performance by providing structured information about entities and their relationships, such as complementary or substitutable relations between products or product types, which can be utilized in recommender systems. However, relation labeling in KGs remains a challenging task due to the dynamic nature of e-commerce domains and the associated cost of human labor. Recently, breakthroughs in Large Language Models (LLMs) have shown surprising results in numerous natural language processing tasks. In this paper, we conduct an empirical study of LLMs for relation labeling in e-commerce KGs, investigating their powerful learning capabilities in natural language and effectiveness in predicting relations between product types with limited labeled data. We evaluate various LLMs, including PaLM and GPT-3.5, on benchmark datasets, demonstrating their ability to achieve competitive performance compared to humans on relation labeling tasks using just 1 to 5 labeled examples per relation. Additionally, we experiment with different prompt engineering techniques to examine their impact on model performance. Our results show that LLMs significantly outperform existing KG completion models in relation labeling for e-commerce KGs and exhibit performance strong enough to replace human labeling.
\end{abstract}



\received{20 February 2007}
\received[revised]{12 March 2009}
\received[accepted]{5 June 2009}

\maketitle

\section{Introduction}
Knowledge Graphs (KGs) have emerged as a powerful tool for representing structured information about entities and their relationships. One of the core tasks of KGs is Knowledge Graph Completion (KGC), which is to predict the relations \cite{cui2021type} that haven't been observed between entities. KGC offers significant benefits in e-commerce, such as relation labeling in product types. By capturing complementary or substitutable relations between product types, KGs enable e-commerce platforms to provide more accurate recommendations for users. However, the process of relation labeling in KGs faces numerous challenges, including the dynamic nature of e-commerce domains and the increasing cost of human labor.

Recent Large Language Models (LLMs), e.g., the 175B-parameter GPT-3 \cite{brown2020language} and the 540B-parameter PaLM \cite{chowdhery2022palm}, are model with a very language amount of parameters. They show surprising abilities, which are called emergent abilities \cite{wei2022emergent}) in solving a series of complex tasks.
A remarkable application of LLMs is ChatGPT \footnote{https://chat.openai.com/}, which adapts the LLMs from the GPT series for dialogue, presents an amazing conversation ability with humans. The powerful capabilities of these models present a potential solution for the challenges faced in relation labeling in e-commerce KGs. LLMs can understand the semantic meanings of product types without training. This paper aims to conduct an empirical study of LLMs for relation labeling in e-commerce KGs, specifically focusing on their few-shot learning capabilities and effectiveness in predicting relations between product types with limited labeled data.

In our experiments, we focus on examining the KGC between product types \cite{lian2008effects} to predict the complementary \cite{ma2021neat, ma2023personalized} and substitutable relations. Product types serve to categorize and group similar products together. While retail platforms like Amazon, eBay, and Walmart may offer millions of distinct products, the number of product types typically remains below 10 thousand. This relatively small number allows for a more nuanced and accurate definition of product relationships. Additionally, product types are well-defined in natural language and can be effectively modeled by KGs, making the research problem well-suited for KG completion. Specifically, given an source (src) product type, our goal is to predict whether it 'is\_complementary\_to', 'is\_substitutable\_for' or 'is\_irrelevant\_to' to another destination (dst) product type. These identified product types can then be utilized to generate high-quality recall item sets for downstream item-level complementary or substitutable recommendations.

In this paper, we evaluate various LLMs, including PaLM \cite{chowdhery2022palm} and GPT-3.5 \cite{brown2020language}, on benchmark datasets to assess their performance on relation labeling tasks using as few as 1 to 5 labeled examples per relation. We also experiment with different prompt engineering techniques to examine their impact on model performance. Our study demonstrates that LLMs significantly outperform existing KG completion models in relation labeling for e-commerce KGs and exhibit performance levels strong enough to replace human labeling. Moreover, LLMs are not only capable of predicting relations but also provide explanations for their labeling decisions regarding the product type pairs in a given relation. Furthermore, we discover that the explanations provided by LLMs is very likely to be agreed by humans if they read them and then change their own labeling results.

This paper is structured as follows: Section 2 describes the settings and datasets used in our experiments. Section 3 presents the results and discussions of the impact of prompt engineering. Section 4 illustrates the labeling results comparison between humans and LLMs. Section 5 shows the comparison experiments between an LLM model PaLM and eight KG models. Section 6 provides an overview of related work in the areas of knowledge graph completion and LLM applications. Finally, Section 7 concludes the paper and suggests future research directions.

Our contributions are summarized as follows:
\begin{itemize}
    \item To the best of our knowledge, this paper represents the first attempt to apply LLMs to KGC tasks in e-commerce contexts. We demonstrate that LLMs possess robust capabilities in predicting complementary and substitutable relations between product types, facilitated by their adeptness at processing natural language.
    
    \item In our experiments, we explore various prompts and identify the most effective way to frame our target task in a few-shot learning context. The performance achieved through our proposed prompt engineering approach is competitive with human labeling and can be readily applied in real-world business scenarios.

    \item We find that LLMs are much powerful than the state-of-the-art KG models with a minimum improvement of 40.6\%. The experiments also demonstrate that LLMs are scalable especially when the number of labeled data is limited.
\end{itemize}


\section{Experiment Settings}
In the experiments of KG relation labeling with LLMs, we first introduce the datasets and their statistics. We considered product types from the Electronics department in Walmart \footnote{walmart.com} and the aisles as product types in online grocery Instacart \footnote{https://www.kaggle.com/c/instacart-market-basket-analysis} \cite{instacart_2017}. The ground truth of the relation labeling is from the consensus of different people through crowdsourcing so we assume this label can be fairly used to evaluate the performance. The temperature of the LLMs are set to 0.0 to ensure the consistent and stable outputs.


In the Electronic dataset, we sampled 1045 pairs of product types, where the labels of the ground truth are 769 for `irrelevant', 264 for `complementary' and 12 for `substitutable'. In the Instacart dataset, we sampled 400 pairs of product types based on their co-occurrence frequency, with 244 `irrelevant' lables, 166 `complementary' labels, and 10 `substitutable' labels.
As each product type is a set of similar products, on the product type level the number of relation `substitutable' is relatively small. 

The LLM's predictions and consensus human labels are evaluated on overall accuracy, precision and recall corresponding to complementary or substitutable labels. For evaluating LLMs and humans, we use human labels as ground truth. The accuracy is calculated as $N_{common\_labels} / N_{total\_labels}$ ,
where $common\_labels$ means the common labels between human and LLM labeling results. 
The precision is calculated for complementary or substitutable relation respectively as $N_{common\_labels}  / N_{LLM\_predicted}$ and the recall is calculated for each relation as $N_{common\_labels} / N_{human\_labeled}$.


\section{Prompt Engineering}
The effectiveness of LLMs in various natural language processing tasks often relies on the design of suitable prompts. In this section, we describe our approach to design prompts for LLMs for the task of relation labeling of product types in e-commerce KGs. We apply PaLM \cite{chowdhery2022palm} and GPT-3.5 \cite{brown2020language} to evaluate its performance on relation labeling in e-commerce.

\begin{figure}[H]
    \centering
    \includegraphics[width=\textwidth]{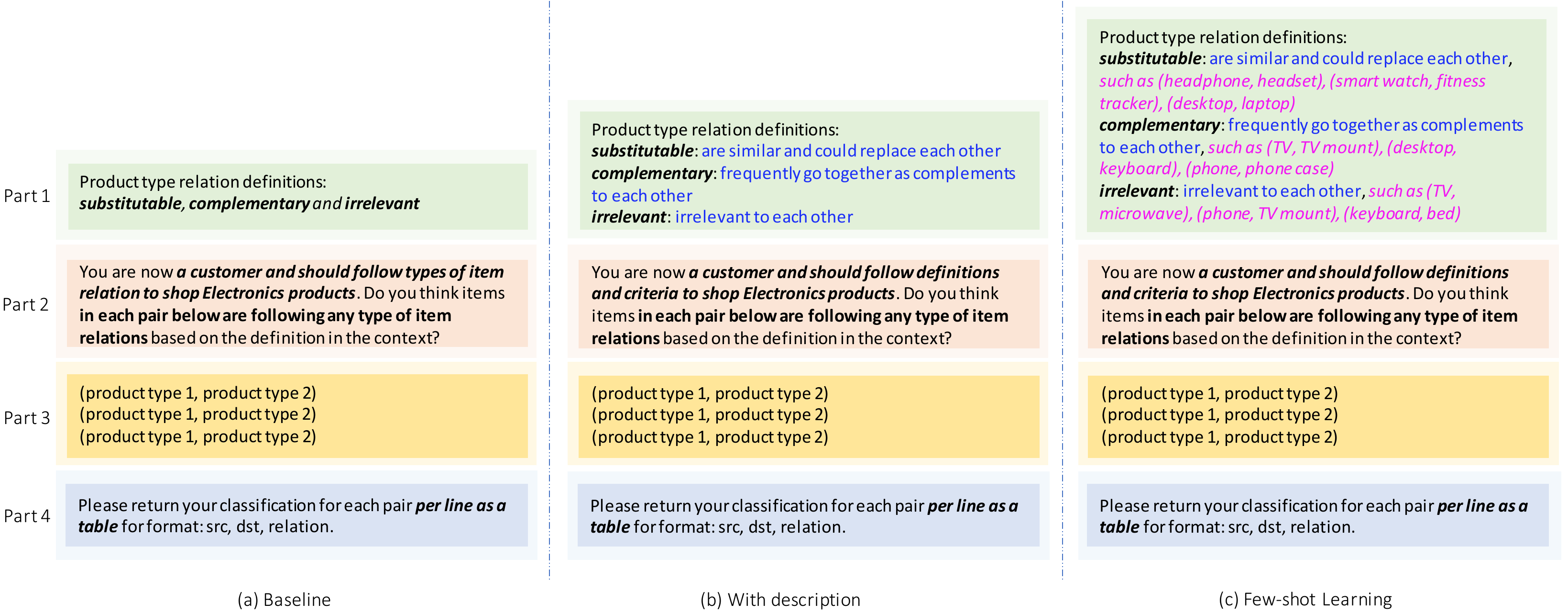}
    \caption{Prompt examples with different principles.}
    \label{fig:prompt}
\end{figure}

To design effective prompts for LLMs, we follow four guiding principles as follows. In Fig. \ref{fig:prompt}, each part in the prompt examples corresponds to a principle.
\begin{itemize}
    \item \textit{Clarity (Part 1)}: Ensure that the prompts can clearly describe \textit{\color{blue}the definition of the relation labeling task}, providing enough context for LLMs to understand the task and the desired output. Few-shot Learning may also be applied as a limited number of examples of the task (e.g., \textit{\color{magenta}pairs of product types in pink} for each item relationship.).

    \item \textit{Relevance (Part 2)}: Set up a role of the LLM and the context of the e-commerce scenario to enhance the model's understanding of the task.
    
    \item \textit{Format (Part 3\&4)}: Frame the input data in the prompts (Part 3) with a clear tuple-like format. The output of the LLM (Part 4) should also follow a certain format to make the results readable.
\end{itemize}

Next, we will introduce how these principles affect the relation labeling performance of LLM. Based on the principles above, we compare the effect of different principles step by step by completing the prompt in Part 1 in Fig. \ref{fig:prompt}.

(a) The baseline. Here we define the baseline prompt with role of LLM, the relation labeling task with a scenario and the output format with Markdown. The accuracy of the baseline prompt is \textbf{0.575}.

(b) With relation description. With the principle Clarity, we also take the part 1 in the Fig.1 into consideration. The prompt will be changed as follows. We highlight the difference in blue. The accuracy of the prompt with relation description is \textbf{0.676}, with a \textbf{17.6\%} improvement compared to the baseline prompt.

(c) With few-shot Learning. On top of the description of relations, we also give each relation a few examples to guide LLMs in performing the task with minimal labeled data. The few-shot examples are highlighted in pink. The accuracy of the prompt with relation description is \textbf{0.738}, with a \textbf{28.3\%} improvement compared to the baseline prompt.

In Table \ref{tab:prompt}, we also put the complete experiment results of prompt engineering on Electronics and Instacart datasets. Please note that we only apply PaLM on Electronics dataset because of data privacy issue. In Instacart dataset, PaLM's results are better than GPT-3.5's for all the prompts in terms of Accuracy. The precision and recall scores for the 'substitutable' relation are relatively low because this relation appears infrequently; thus, the results are easily affected by incorrect predictions, leading to a bias in the scores. From the results of these two tables, we can find that the relation definition and few-shot learning with 3 or 5 examples can lead to a significant improvement on the prediction accuracy.

\begin{table}[H]
    \centering
    \begin{tabular}{l|l|l|ll|ll|l}
    \toprule
           \multirow{2}{*}{\textbf{Dataset}} & \multirow{2}{*}{\textbf{LLM}} &  \multirow{2}{*}{\textbf{Prompt}} & \multicolumn{2}{c}{\textbf{Complementary}} & \multicolumn{2}{c}{\textbf{Substitutable}} &   \multirow{2}{*}{Accuracy} \\
          &  &       &     Precision &     Recall &     Precision &     Recall &   \\
    \midrule
        \multirow{5}{*}{\textbf{Electronics}} & \multirow{5}{*}{PaLM} &    Baseline &         0.389 &      \textbf{0.807} &         0.083 &        0.500 &      0.575 \\
        &   &       zero\_shot &         0.424 &      0.678 &          \textbf{0.240} &        0.500 &      0.676 \\
         &  &     one\_shot &         0.446 &      0.667 &         0.227 &      0.417 &      0.695 \\
    &   &few\_shot\_3 &         0.506 &      0.633 &         0.222 &        0.500 &      \textbf{0.738} \\
    &   &few\_shot\_5&         \textbf{0.507} &       0.580 &         0.136 &        0.500 &      0.725 \\
    \midrule
        \multirow{10}{*}{\textbf{Instacart}} & \multirow{5}{*}{PaLM} &   Baseline &         0.599 &      \textbf{0.786} &         0.167 &      0.444 &      0.645 \\
            &   &  zero\_shot &         0.705 &      0.656 &         0.161 &      \textbf{0.556} &      0.699 \\
            &   &   one\_shot &         0.664 &       0.740 &           \textbf{0.300} &      0.333 &      0.712 \\
            &   & few\_shot\_3 &         0.699 &      0.725 &         0.222 &      0.444 &      0.726 \\
            &   & few\_shot\_5 &         \textbf{0.711} &      0.733 &          0.250 &      0.444 &      \textbf{0.739} \\
    \cmidrule{2-8}
            & \multirow{5}{*}{GPT-3.5} &   Baseline &         0.636 &      0.519 &         0.091 &      \textbf{0.778} &      0.572 \\
            &   &  zero\_shot &         0.595 &      0.695 &         0.125 &      0.444 &      0.632 \\
            &   &   one\_shot &         0.598 &      0.656 &         0.135 &      0.556 &      0.622 \\
            &   & few\_shot\_3 &         \textbf{0.659} &      0.618 &         0.133 &      0.667 &      0.635 \\
            &   & few\_shot\_5 &         0.632 &      \textbf{0.695} &         \textbf{0.167} &      0.444 &      \textbf{0.666} \\
    \bottomrule
    \end{tabular}    
    \caption{LLM label results on Electronics and Instacart datasets. (1) PaLM label results on Electronics product types. (2) PaLM and ChatGPT results on Instacart product types.}
    \label{tab:prompt}
\end{table}

\section{LLM as Individual Human Labeler}

Human consensus results offer costly but accurate labels through crowdsourcing, enhancing the labeling quality by incorporating the input of multiple labelers. Although individual labelers may make mistakes in labeling tasks, they can still provide different yet valid labels compared to consensus results due to their diverse backgrounds and experiences. For instance, if two labelers come from different regions with distinct dietary habits, they might offer different but valid labels of relationships for grocery product types.
To further investigate the LLM's performance and the gap between it and individual labelers, we compare LLM's results with those of individual human labelers under \textbf{independent labeling} and \textbf{dependent labeling} settings. For all subsequent experiments, we consider the results from PaLM with the prompt (few\_shot\_5) as LLM's labels for Instacart product type pairs, and the results from PaLM with the prompt (few\_shot\_3) as LLM's labels for Electronics product type pairs, due to their superior performance on human consensus results.


\subsection{LLM Results vs Individual Human Labelers (independent labeling)}

In this experiment, two human labelers with different cultural backgrounds but extensive experience in e-commerce shopping independently label the relationships of all pairs of the aforementioned Electronics and Instacart product types, respectively. They are not allowed to discuss their findings with each other or review the LLM's results.
To initially understand the gap between the LLM and individual labelers, we treat each individual human labeler's results as ground truth and evaluate the LLM's results with those of the two individual labelers, respectively. To further understand the impact of a human labeler's background on labeling tasks, we treat labeler 2's results as ground truth and evaluate labeler 1's results. We report the precision, recall, and accuracy metrics as defined in Section 2 in Table~\ref{tab:double_blind}. It is important to note that there is no actual ground truth between the two individual human labelers; the precision and recall reported here simply represent the proportion of agreed-upon labels for human labelers 1 and 2.

From the accuracy results in Table~\ref{tab:double_blind}, for Electronics product types, the human-human accuracy is 0.76, and the LLM's accuracy with labeler 2 is very close to the human-human results. On the Instacart dataset, the accuracy between the LLM and labeler 2 (0.665) surpasses the accuracy between human and human (0.598). Notably, from the precision and recall results, LLM usually has a low precision value for 'substitutable' labels with both human labelers, which indicates LLM tends to generate more 'substitutable' results.
In summary, LLM's performance is comparable to that of individual labelers, considering the accuracy between LLM and human as well as the accuracy between human and human. Furthermore, an individual's background does indeed influence their labeling performance.

\begin{table}[H]
    \centering
    \begin{tabular}{l|ll|ll|ll|l}
    \toprule
            \multirow{2}{*}{\textbf{Dataset}} & \multirow{2}{*}{\textbf{Prediction}} &  \multirow{2}{*}{\textbf{Ground truth}} & \multicolumn{2}{c}{\textbf{Complementary}} & \multicolumn{2}{c}{\textbf{Substitutable}} &        \multirow{2}{*}{Accuracy} \\
         &   &     &     Precision &     Recall &     Precision &     Recall &    \\
    \midrule
    \multirow{3}{*}{Electronics} &        LLM &  Labeler 1 &      \textbf{0.822} &   0.557 &       0.161 &   \textbf{0.833} &   0.687 \\
             &        LLM &  Labeler 2 &      0.742 &   0.652 &      0.322 &   0.769 &       0.74 \\
             &  Labeler 1 &  Labeler 2 &      0.651 &   \textbf{0.843} &      \textbf{0.667} &   0.308 &       \textbf{0.76} \\
    \midrule
      \multirow{3}{*}{Instacart} &        LLM &  Labeler 1 &       0.356 &   0.638 &      0.0968 &   0.214 &       0.547 \\
             &        LLM &  Labeler 2 &      \textbf{0.654} &   \textbf{0.687} &      0.129 &   \textbf{0.308} &     \textbf{0.665} \\
             &  Labeler 1 &  Labeler 2 &      0.628 &   0.369 &      \textbf{0.286} &   0.308 &     0.598 \\
    \bottomrule
    \end{tabular}
    \caption{Evaluation with human independently labeled results.}
    \label{tab:double_blind}
\end{table}

\begin{table}[H]
    \centering
    \begin{tabular}{l|ll|ll|ll|l}
    \toprule
            \multirow{2}{*}{\textbf{Dataset}} & \multirow{2}{*}{\textbf{Prediction}} &  \multirow{2}{*}{\textbf{Ground Truth}} & \multicolumn{2}{c}{\textbf{Complementary}} & \multicolumn{2}{c}{\textbf{Substitutable}} &        \multirow{2}{*}{Accuracy} \\
         &   &     &     Precision &     Recall &     Precision &     Recall &    \\
    \midrule
            \multirow{3}{*}{Electronics} &        LLM &  Labeler 1 &         \textbf{0.832} &      0.622 &          0.29 &        \textbf{0.9} &       0.74 \\
            &        LLM &  Labeler 2 &         0.782 &      0.675 &         0.387 &        0.8 &      0.767 \\
            &  Labeler 1 &  Labeler 2 &         0.696 &      \textbf{0.803} &           \textbf{0.8} &      0.533 &       \textbf{0.78} \\
    \midrule
            \multirow{3}{*}{Instacart} &        LLM &  Labeler 1 &         \textbf{0.723} &      \textbf{0.932} &         \textbf{0.548} &      \textbf{0.944} &       \textbf{0.82} \\
            &        LLM &  Labeler 2 &         0.669 &      0.717 &         0.161 &      0.417 &      0.695 \\
            &  Labeler 1 &  Labeler 2 &         0.712 &      0.601 &         0.278 &      0.417 &      0.718 \\
    \bottomrule
    \end{tabular}
    \caption{Human relabeling based on LLM results. LLM label results are generated \textbf{with explanations}.}
    \label{tab:single_blind}
\end{table}


\subsection{Human Relabeling based on LLM Results (dependent labeling)}


Owing to the limitations of individual knowledge, a human labeler may lack information for some products, resulting in incorrect labeling outcomes. To further assess the quality of LLM's labels, we modify part 4 of our prompt template to request that LLM provide both labels and explanations. We then ask our two labelers to re-label the product types for both datasets, taking into account LLM's labels and explanations. The evaluation results for LLM-human and human-human comparisons are presented in Table~\ref{tab:single_blind}.


On the Electronics dataset, the accuracy between both LLM-human and human-human comparisons is slightly improved compared to the results in Table~\ref{tab:double_blind}. However, on the Instacart dataset, the accuracy between LLM-labeler 1 and human-human increased by more than 10\%. This could be due to the fact that relations between most electronic products are objective and easy to determine, while relationships between grocery products are more subjective and challenging to ascertain. For example, people with different dietary habits might have differing opinions on grocery product relationships. Additionally, the agreement between LLM-labeler 1 on the Instacart dataset is much higher than the agreement between human-human, indicating that LLM's explanations have convinced labeler 1 in many grocery pair cases.

Typically, we notice that labeler 1 changed 110 labels after seeing LLM's labels and explanations compared with the independent labeling task, mostly due to 62 pairs changed from `irrelevant' to `complementary' and 25 from `complementary' to `irrelevant'. 
For example, when the src product type is `yogurt' and the dst product type is `fresh dips tapenades', labeler 1 tags them as `irrelevant' mainly because they are not common combination in the food culture of labler 1. LLM tags them as `complementary' with explanation `\textit{\color{purple}yogurt and fresh dips tapenades can both be used as a snack or appetizer. they can also be eaten separately}', which convinces labeler 1 to change the labels. 
Another examples `canned jarred vegetables' and `milk'. Labeler 1 change label from `complementary' to `irrelevant' after checking LLM's explanation `\textit{\color{purple}canned and jarred vegetables are both processed forms of vegetables, while milk is a dairy product. they are not typically used in the same recipes, and they do not have the same nutritional value}', which addresses more on nutrition compatibility.

Through both independent and dependent labeling tasks, we demonstrate that LLMs can perform competitively compared to human labelers, taking into account individual differences. Additionally, LLMs provide valuable labeling explanations that contribute to better label quality in KG completion tasks.

\section{Comparison Experiments}
\begin{figure}[H]
    \begin{subfigure}[]{0.45\textwidth}
        \centering
        \includegraphics[height=2.2in]{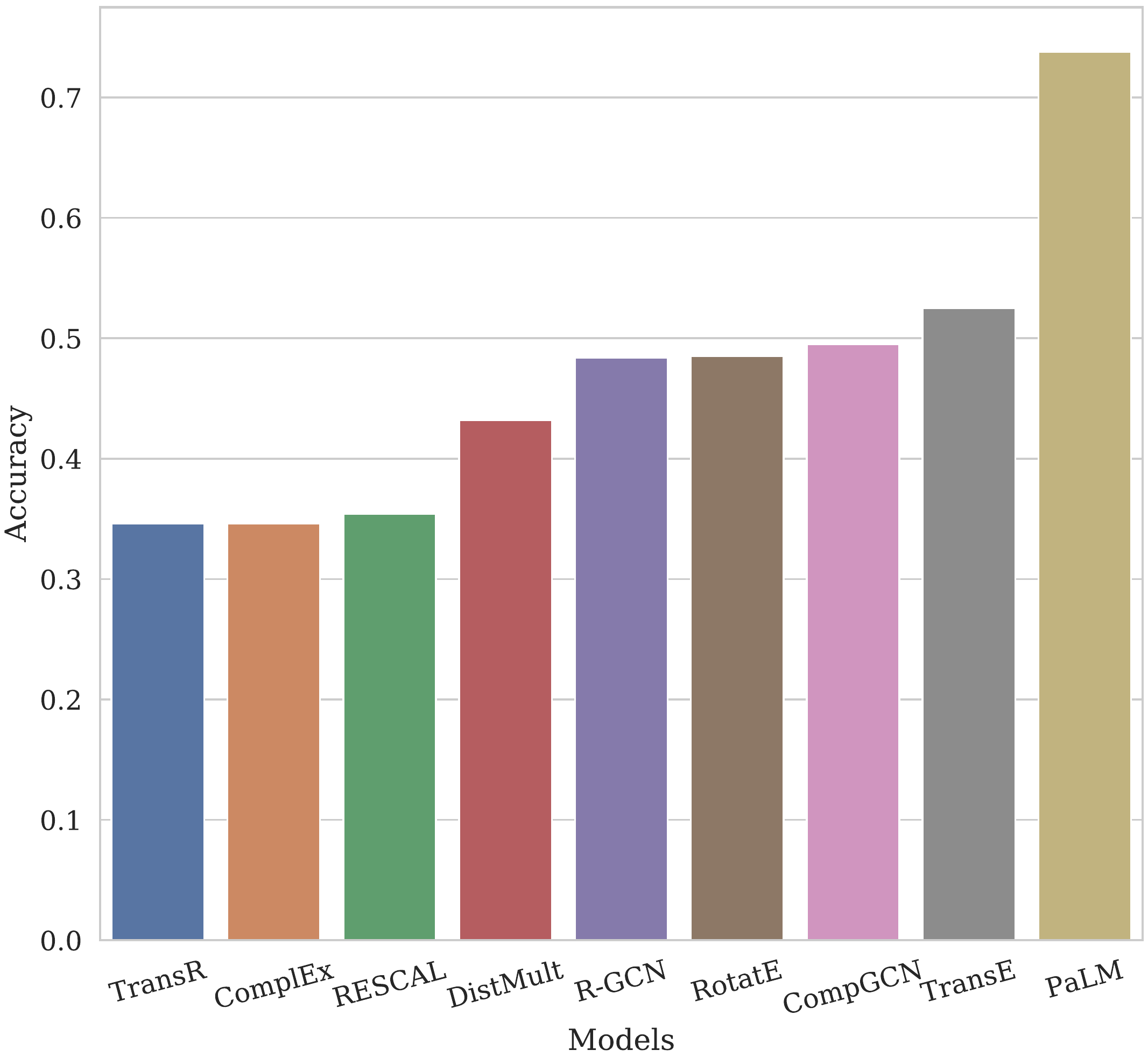}
        \caption{Electornics} 
    \end{subfigure}
    \hspace{0.001\textwidth}
    \begin{subfigure}[]{0.45\textwidth}
        \centering       
        \includegraphics[height=2.2in]{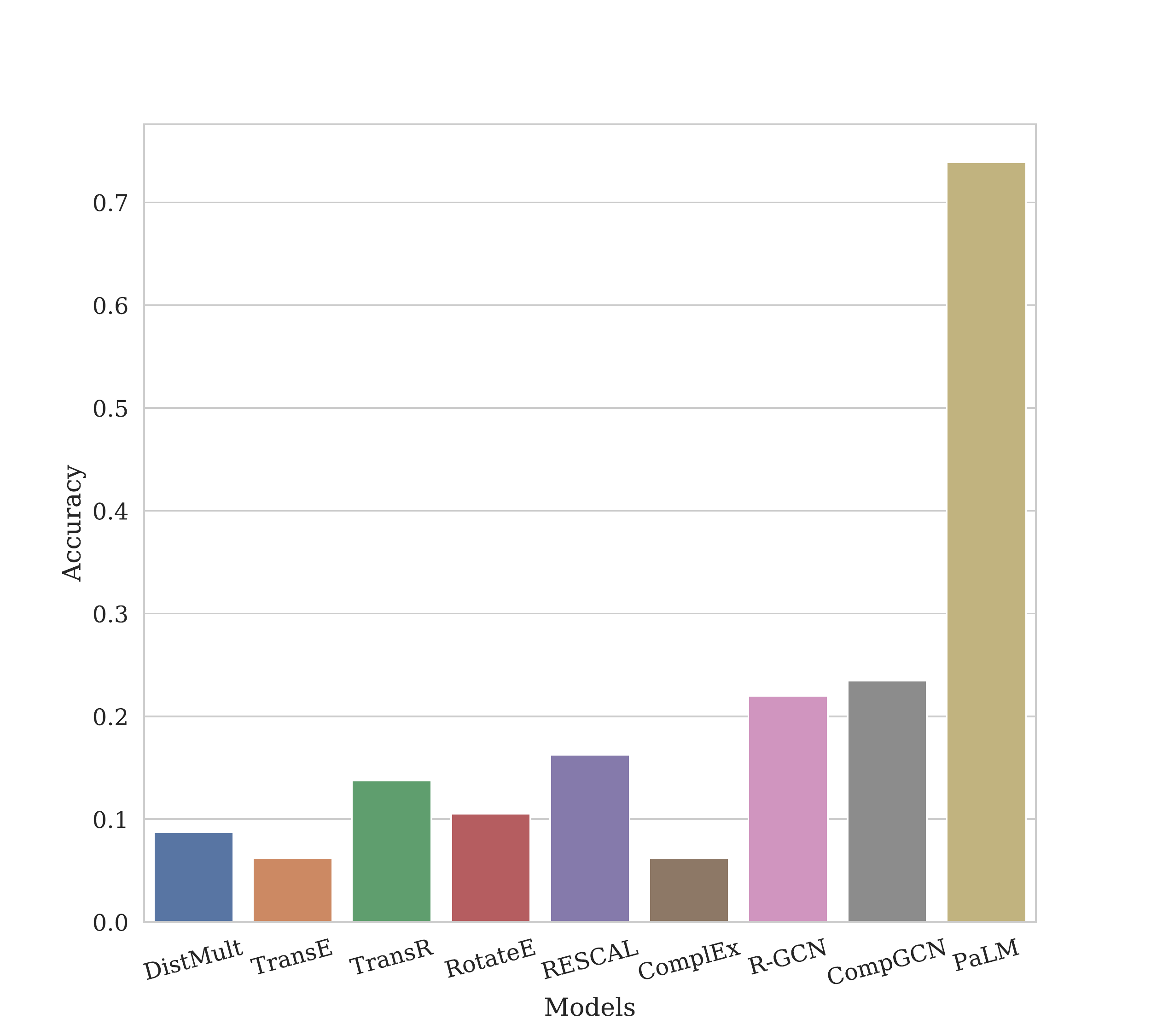}        
        \caption{Instacart} 
        
    \end{subfigure}
    \caption{Comparision of accuracy between KG models and the LLM in the  Electornics and Instacart dataset.} 
    \label{fig:exp}
\end{figure}

We conduct comparison experiment of an LLM PaLM with different KG models. The baseline models are TransE \cite{bordes2013translating}, TransR \cite{lin2015learning}, DistMult \cite{yang2015embedding}, ComplEx \cite{trouillon2016complex}, RESCAL \cite{nickel2011three}, R-GCN \cite{schlichtkrull2018modeling} and CompGCN \cite{vashishth2019composition}. The product types in all KG models are initialized with word embeddings from Word2Vec \cite{mikolov2013distributed}. The experiments are conducted on both Electronics and Instacart datasets with human consensus labels, and we split the datasets as 80\% for training, 10\% for validation and 10\% for testing. The detailed results of the experiments are shown in Fig. \ref{fig:exp}. From the two figures, we have the following observations:
\begin{itemize}
    \item We observe that PaLM significantly outperforms all knowledge graph models on both datasets, with the minimum improvement being 40.6\%. This can be attributed to the fact that KG models require a substantial amount of training data, while relation labeling in e-commerce is expensive, resulting in limited labeled data for our task. Furthermore, the labeled data does not cover all product types, leading to instances where some product types never appear in the training set. In contrast, LLMs leverage their understanding of human language to enhance the accuracy of their predictions, even when dealing with limited training data or unseen product types.

    \item The LLM model PaLM exhibits similar accuracy on both datasets. However, in Fig. \ref{fig:exp} (b), KG models perform poorly on the Instacart dataset due to the limited availability of only 320 training pairs. This observation highlights that, in contrast to KG models that require large amounts of data for parameter optimization, LLM models are more scalable and their performance is less influenced by the quantity of labeled data.
\end{itemize}

\section{Related Works}
\subsection{Knowledge Graph Completion in E-commerce}
Knowledge Graphs (KGs) have garnered significant attention in recent years due to their ability to represent structured information about entities and their relationships. They have been widely adopted across various domains, including e-commerce, to enhance user experiences and facilitate decision-making. In this section, we review the literature on Knowledge Graph Completion (KGC) in the context of e-commerce.

With the emergence of embedding techniques, several KGC approaches have employed embeddings to represent entities and relations in e-commerce KGs. These methods, such as TransE \cite{bordes2013translating}, DistMult \cite{yang2015embedding}, and ComplEx \cite{trouillon2016complex}, learn low-dimensional vector representations of entities and relations, enabling the discovery of complex patterns and relationships within the KG. Recent advances in neural networks have led to the development of more sophisticated KGC methods in e-commerce. Convolutional Neural Networks (CNNs) have been employed for KGC tasks, such as in ConvE~\cite{dettmers2018convolutional}, where the model learns embeddings by exploiting local and global connectivity patterns in the graph. Similarly, Graph Neural Networks (GNNs) \cite{kipf2016semi} have demonstrated its capacities to capture both structural and semantic information~\cite{wang2019kgat, li2020dynamic, chen2022grease, li2022time, liu2020heterogeneous, liu2021medical}. They are also applied to KGs, such as R-GCN~\cite{schlichtkrull2018modeling} and CompGCN~\cite{vashishth2019composition}. The few-shot learning in KGC \cite{zhang2020few, xiong2018one} can also improve the performance when labeled data is scarce. KGC methods have a significant impact on many applications in e-commerce, including recommender systems \cite{wang2019kgat, liu2020basket, li2021pre}, product relation labeling \cite{xu2020product} and product taxonomy \cite{martel2021taxonomy}.

While these methods have demonstrated improved performance, they still exhibit limitations in comprehending natural language, which is crucial for e-commerce KGs. Moreover, these approaches continue to face challenges in addressing the scarcity of labeled data, primarily due to the expensive cost of human labor. By employing LLMs, we can capitalize on their capacity to understand natural language and label relations within the context of few-shot learning, potentially overcoming these challenges and enhancing KG completion accuracy in e-commerce domains.

\subsection{LLM Applications in E-commerce}
Large Language Models (LLMs) have gained significant traction in recent years due to their remarkable performance in a wide range of natural language processing related tasks \cite{wang2023chatgpt, dai2023chataug, luo2023chatgpt}. In this section, we review the literature on LLM applications in e-commerce.

One of the most common applications of LLMs in e-commerce is the enhancement of recommender systems. By leveraging LLMs' natural language understanding capabilities, researchers have been able to provide more accurate and personalized product recommendations for users. For example, LLMs have been used to  learn from users' behaviors in natural language so that they can serve as recommender systems to directly make recommendations \cite{cui2022m6, geng2022recommendation}. Moreover, as LLMs' success in conversaional AI, there are some new applications such as conversational recommendation \cite{wang2021finetuning} to enhance the customers' experience.

LLMs have been employed across a range of applications in e-commerce, including customer support \cite{george2023review}, sentiment analysis \cite{wang2023chatgpt}, and text classification \cite{kant2018practical}. These applications help e-commerce platforms better understand product information, customer feedback, and preferences, ultimately leading to more targeted marketing strategies and improved user experiences. However, their application in Knowledge Graph Completion (KGC) remains relatively unexplored, particularly in the context of e-commerce. In this paper, we aim to bridge this gap by investigating LLMs' potential for predicting complementary and substitutable relations between product types in e-commerce KGs.
\section{Conclusion}

This study contributes to the understanding of LLMs' potential in e-commerce KG completion tasks and demonstrates their value in overcoming challenges associated with limited labeled data and human labor costs. Our results revealed that LLMs significantly outperform existing KG completion models in relation labeling for e-commerce KGs and exhibit performance strong enough to replace human labeling. As a pioneering effort in applying LLMs to KGC tasks in e-commerce, our findings pave the way for future research and practical applications of LLMs in e-commerce such as item description summarizing or recommendation.

\bibliographystyle{ACM-Reference-Format}
\bibliography{sample-base}


\end{document}